\documentclass[12pt,twoside]{article}
\usepackage[mathscr]{eucal}
\usepackage{amsmath,amsfonts,amssymb,amsthm,mathabx,empheq}
\usepackage{times}
\usepackage{pdfsync}
\usepackage{cite}
\usepackage{url}
\usepackage{hyperref}
\usepackage{tensor}
\usepackage{color}
\usepackage{multicol}
\usepackage{bbold}

\advance\textheight\topskip
\allowdisplaybreaks[1]

\newcommand\td{\text{d}}
\newcommand\cO{{\cal O}}
\newcommand{\p}{\partial}
\newcommand{\be}{\begin{equation}}
\newcommand{\ee}{\end{equation}}

\def\bz{\bar z}
\def\ga{{\bar\gamma}_{AB}}
\def\ii{\rm i}
\def\cb {}

\newcommand*\xbar[1]{%
  \hbox{%
    \vbox{%
      \hrule height 0.5pt 
      \kern0.3ex
      \hbox{%
        \kern-0.0em
        \ensuremath{#1}%
        \kern-0.0em
      }%
    }%
  }%
}

\DeclareFontFamily{OT1}{rsfs}{} \DeclareFontShape{OT1}{rsfs}{m}{n}{
<-7> rsfs5 <7-10> rsfs7 <10-> rsfs10}{}
\DeclareMathAlphabet{\mycal}{OT1}{rsfs}{m}{n}

\begin{document}
\title{Hawking radiation of Dirac particles from soft-hairy black {\cb holes}}

\author{Wen-Jie Zhang, Pu-Jian Mao and Jun-Bao Wu}

\date{}

\def\mytitle{Hawking radiation of Dirac particles from soft-hairy black {\cb holes}}

\begin{centering}

  \vspace{1cm}

  \textbf{\Large{\mytitle}}

  \vspace{1.5cm}

  {\large Wen-Jie Zhang, Pu-Jian Mao and Jun-Bao Wu }

\vspace{.5cm}

\vspace{.5cm}
\begin{minipage}{.9\textwidth}\small \it  \begin{center}
     Center for Joint Quantum Studies and Department of Physics,\\
     School of Science, Tianjin University, 135 Yaguan Road, Tianjin 300350, China
 \end{center}

\end{minipage}

\end{centering}

\begin{center}
Emails:zhangwenjie19@tju.edu.cn,~pjmao@tju.edu.cn,~junbao.wu@tju.edu.cn
\end{center}

\vspace{1cm}

\begin{center}
\begin{minipage}{.9\textwidth}
  \textsc{Abstract}. In this paper, we study the Hawking radiation of Dirac particles via tunneling formalism from linearly supertranslated Schwarzschild black {\cb holes}. We find that the radiation spectrum and the Hawking temperature remain the same as the one without soft hair.
 \end{minipage}
\end{center}
\thispagestyle{empty}


\section{Introduction}

Black holes are shown to carry soft hairs \cite{Hawking:2016msc,Compere:2016hzt,Afshar:2016wfy,Mao:2016pwq,Grumiller:2016kcp,Hawking:2016sgy,Chu:2018tzu,Choi:2018oel,Choi:2019fuq,Takeuchi:2021rrq}. Consequently, the vacuum of quantum gravity is not unique and a stationary black hole is physically inequivalent to the one with soft hairs. During the course of Hawking evaporation, soft hairs should be radiated through infinity and the assumption of charge {\cb conservation} will induce correlations between the early and late time Hawking radiation which are normally not considered in the semiclassical computation. Moreover, the normal (hard) radiated particles are always accompanied with an infinite cloud of correlated soft particles and the quantum purity can be just restored in principle by correlations between the hard and soft radiated particles \cite{Strominger:2017aeh}. The whole picture makes the soft hair at the position of a possible resolution of the black hole information paradox, see, e.g., \cite{Strominger:2017zoo} for a review and references therein. Meanwhile, infalling observers will automatically observe the soft radiated particles outside the horizon. Taking into account the soft dynamics, it is argued that an infalling observer will never really meet a firewall before reaching the singularity \cite{Pasterski:2020xvn} which indicates that the soft hair can also play an essential role in dissolving the firewall paradox.

A natural question arises {\cb about} how soft hairs can be observed, in particular macroscopically. Recent studies show that soft hairs in three dimensions are purely microscopic \cite{Afshar:2016uax,Sheikh-Jabbari:2016npa}. In four dimensions, the same feature has been verified by checking the Hawking radiation of soft-hairy black holes \cite{Chu:2018tzu,Javadinezhad:2018urv,Compere:2019rof,Lin:2020gva,Takeuchi:2021ibg}. More precisely, the Hawking fluxes from {\cb the} Schwarzschild black hole {\cb are} unaffected by their linearized supertranslation hair \cite{Chu:2018tzu}. The spectrum of Hawking radiation of the Schwarzschild black hole is unchanged after including the dressing of asymptotic states with clouds of soft photons and soft gravitons \cite{Javadinezhad:2018urv}. Unruh effect is invariant under supertranslations \cite{Compere:2019rof}. The Hawking fluxes from Schwarzschild with non-linear supertranslation hair are still the same as the Schwarzschild's one \cite{Lin:2020gva} (see also \cite{Takeuchi:2021ibg}). However counter examples are presented in \cite{Chu:2018tzu,Chiang:2020lem,Wen:2021ahw}. In particular, the Hawking temperature of dynamical black holes is found to be modified by soft hair \cite{Chu:2018tzu,Chiang:2020lem}. Complementary to these investigations where only scalar particles are involved, we study in this paper the Hawking radiation of Dirac particles from soft-hairy black {\cb holes} using the tunneling method. The supertranslation is characterized by an arbitrary function of the angular variables. The resulting solution after a supertranslation will be in a rotating frame. The main motivation of the present work of investigating fermions emitted from a supertranslated black hole is to check if there is correlation between the fermion spin and the supertranslation parameters in the tunneling probability.

The tunneling method proposed by Parikh and Wilczek \cite{Parikh:1999mf} is a very effective way of studying Hawking radiation. In general, the tunneling method involves calculating the imaginary part of the action for the classically forbidden path to evaluate the radiation spectrum. Applying the tunneling method, we examine the Hawking radiation of Dirac particles from Schwarzschild black {\cb holes} with linear supertranslation hairs. We find that the radiation spectrum and the Hawking temperature are unchanged after soft hair implant on black {\cb holes}. When applying the tunneling method for Dirac particles, one needs to properly take care of the angular-dependent part of the action \cite{Kerner:2007rr,Li:2008zra,DiCriscienzo:2008dm,Chen:2008vi,Kerner:2008qv,Chen:2008ge,Jiang:2008gq,Vanzo:2011wq,Chen:2014xgj}. The simplest assumption that solving only the radial equation at a fixed angle may not  {\cb lead to} a solution of the Dirac equation. We find that the linear supertranslation hair dependence in the Hawking spectrum is precisely canceled by taking account of the contribution from the angular-dependent part. Our result is in consistent with the scalar particle case \cite{Chu:2018tzu} and furnishes an important confirmation of the microscopic nature of soft hairs in four dimensions.

\section{Schwarzschild black {\cb holes} with supertranslation hair}

The line element of the supertranslated Schwarzschild black hole in the advanced Bondi coordinates reads \cite{Hawking:2016sgy}
\begin{multline}\label{sch-super}
\td s^2=-\left(V-\frac{M D^2f}{r^2}\right) \td v^2 + 2 \td v \td r - D_A \left(2 V f + D^2 f \right) \td v \td \Theta^A \\
+\left(r^2 \ga + 2 r D_A D_B f - r \ga D^2 f \right) \td \Theta^A \td \Theta^B \ , \quad V=1-\frac{2M}{r} \ ,
\end{multline}
where $\ga$ is the metric of the unit two sphere on the boundary and $D_A$ is the covariant derivative with respect to $\ga$, the index in capital Latin character will be raised and lowered by ${\bar\gamma}^{AB}$ and $\ga$, $D^2=D_A D^A$, $\Theta^A=(z, \bz)$ are the angular coordinates, $f$ is an arbitrary function of the angular variables that characterizes the supertranslation. Up to order $f$ term, $g_{vv}=-\frac{1}{r^2}(r-r_h)(r+r_f)$ where $r_f=\frac12 D^2 f$ and  $r_h=2M + \frac12 D^2 f$ is the location of event horizon. Since the supertranslation is infinitesimal, one can drop the terms with order higher than $\cO(f)$ \cite{Hawking:2016sgy}. Physically, the solution can be considered as throwing in an asymmetric null shock into the Schwarzschild black hole. So this solution is physically inequivalent to the Schwarzschild black hole.

There are two different approaches to compute the imaginary part of the action for the emitted particle, namely the null geodesic method used by Parikh and Wilczek \cite{Parikh:1999mf} and the Hamilton-Jacobi method proposed by Agheben et. al. \cite{Angheben:2005rm}. The former relies on a specific choice of coordinates while the later can apply to any coordinate system that is regular at the horizon. Moreover, the Hamilton-Jacobi equation can be emerged from a WKB approximation of the Klein-Gordon equation \cite{Kerner:2007rr}. For a scalar field, assuming that
\be
\phi=\text{exp}\left[\frac{{\cb \ii}}{\hbar} I + I_1 + \cO(\hbar)\right]\ ,
\ee
one can obtain the usual Hamilton-Jacobi equation from the lowest order in $\hbar$ expansion of the Klein-Gordon equation when identifying $I$ as the action of the tunneling particle. Such a treatment can be further applied to Dirac equation which allows one to study the tunneling of {\cb fermions} \cite{Kerner:2007rr}. We consider massive spinor field $\Psi$ with mass $\mu$. The covariant Dirac equation reads
\be\label{Dirac}
{\cb \ii} \hbar \gamma^a e_a^\mu \nabla_\mu \Psi - \mu \Psi = 0 \ ,
\ee
where $\nabla_\mu$ is the spinor-covariant derivative defined by $\nabla_\mu=\p_\mu + \frac14 {\omega^{ab}}_\mu \gamma_{[a}\gamma_{b]}$ and ${\omega^{ab}}_\mu$ is the spin connection. The co-tetrad fields $e^a_\mu$ according to the metric \eqref{sch-super} are chosen as\footnote{Notice the choice of $\eta_{ab}$ in eq.~\eqref{eta}.}
\begin{align}
&e^0_\mu=\left[-1,0,0,0\right] \ ,\\
&e^1_\mu=\left[ \frac12 g_{vv}, 1 ,  g_{vA}\right] \ ,\\
&e^2_\mu=\left[0,0,f_1(x^\mu),f_2(x^\mu)\right] \ ,\\
&e^3_\mu=\left[0,0,0,f_3(x^\mu)\right] \ ,
\end{align}
where $f_1=\sqrt{g_{zz}},\,f_2=\frac{g_{z\bz}}{\sqrt{g_{zz}}},\,f_3=\sqrt{g_{\bz\bz} - \frac{g_{z\bz}^2}{g_{zz}}}$.
We will consider the $\gamma$ matrices as
\begin{align}
&\gamma^0=\frac{1}{\sqrt{2}}\begin{pmatrix}
{\cb\ii} & 0 \\ 0 & - {\cb\ii}
\end{pmatrix}  + \frac{1}{\sqrt{2}} \begin{pmatrix}
0 & \sigma^3 \\ \sigma^3 & 0
\end{pmatrix}\ ,\quad \gamma^2=\begin{pmatrix}
0 & \sigma^1 \\ \sigma^1 & 0
\end{pmatrix}\ ,\\
&\gamma^1=\frac{1}{\sqrt{2}}\begin{pmatrix}
{\cb\ii} & 0 \\ 0 & - {\cb\ii}
\end{pmatrix}  - \frac{1}{\sqrt{2}} \begin{pmatrix}
0 & \sigma^3 \\ \sigma^3 & 0
\end{pmatrix} \ ,
\quad
\gamma^3=\begin{pmatrix}
0 & \sigma^2 \\ \sigma^2 & 0
\end{pmatrix}\ ,
\end{align}
where the matrices $\sigma^k(k = 1,2,3)$ are the Pauli matrices. The $\gamma^a$ matrices satisfy the anti-commutation relations $\{\gamma^a,\gamma^b\}=2\eta^{ab}\times \mathbb{1}$ where
\be
\eta_{ab}=\eta^{ab}=\begin{pmatrix}
0 & -1 & 0 & 0  \\ -1 & 0 & 0 & 0  \\ 0 & 0 & 1 & 0  \\ 0 & 0 & 0 & 1
\end{pmatrix}\ .\label{eta}
\ee
The spin-up and spin-down spinor fields are assumed to have the form \cite{Kerner:2007rr}
\be
\Psi_{\uparrow}=\begin{pmatrix}
A(x^\mu)  \\  0  \\ B(x^\mu)   \\ 0
\end{pmatrix}\text{exp}\left[\frac{{\cb \ii}}{\hbar}I_{\uparrow}(x^\mu)\right] \ ,\quad
\Psi_{\downarrow}=\begin{pmatrix}
0 \\ C(x^\mu)  \\  0  \\ D(x^\mu)
\end{pmatrix}\text{exp}\left[\frac{{\cb \ii}}{\hbar}I_{\downarrow}(x^\mu)\right] \ .
\ee
We will only demonstrate the spin-up case explicitly while the spin-down case is fully in analogue.

Inserting the spin-up ansatz into the covariant Dirac equation \eqref{Dirac} and applying the WKB approximation, one can arrive at the following four equations at the leading order in the expansion of $\hbar$:
\begin{align}
&\sqrt{2} B \left[\p_r I (g_{vv} - 2) - 2 \p_v I \right] + A \left[4 \mu + \sqrt2 {\cb\ii} (2 + g_{vv}) \p_r I - 2\sqrt2 {\cb\ii} \p_v I \right] = 0 \ ,\\
&B \left[ (f_2 + {\cb\ii} f_3 ) (\p_z I - g_{vz} \p_r I) - f_1 (\p_{\bz} I - g_{v\bz} \p_r I ) \right]= 0 \ ,\label{2}\\
&\sqrt{2} A \left[\p_r I (g_{vv} - 2) - 2 \p_v I \right] + B \left[4 \mu - \sqrt2 {\cb \ii} (2 + g_{vv}) \p_r I + 2\sqrt2 {\cb\ii} \p_v I \right] = 0 \ ,\\
&A \left[ (f_2 + {\cb\ii} f_3 ) (\p_z I - g_{vz} \p_r I) - f_1 (\p_{\bz} I - g_{v\bz} \p_r I ) \right]= 0 \ .\label{4}
\end{align}
For the first and third equations, the non-trivial solution for A and B exists if and only if the determinant of the coefficient matrix vanishes. Then we obtain
\be\label{r}
g_{vv} (\p_r I)^2 - 2 \p_v I \p_r I + \mu^2= 0 \ .
\ee
Equations \eqref{2} and \eqref{4} yield the same equation regardless of $A$ or $B$, i.e.,
\be\label{angle}
(f_2 + {\cb\ii} f_3 ) (\p_z I - g_{vz} \p_r I) - f_1 (\p_{\bz} I - g_{v\bz} \p_r I )=0 \ .
\ee
Because $\frac{\p}{\p v}$ is a Killing vector in the supertranslated spacetime, we can separate the variables for $I(v,r,z,\bz)$ as follows:
\be
I =- \omega v + R(r,z,\bz) + K \ ,
\ee
where $\omega$ is the fermionic particle's energy and $K$ is a complex constant. Then \eqref{r} can be solved out as
\begin{align}
&R_+=\int \td r \, \frac{\omega + \sqrt{\omega^2 - \mu^2 g_{vv}}}{-g_{vv}} + \Phi_+(z,\bz) \ ,\label{+} \\
&R_-=\int \td r \, \frac{\omega - \sqrt{\omega^2 - \mu^2 g_{vv}}}{-g_{vv}} + \Phi_-(z,\bz) \ ,\label{-}
\end{align}
where $R_+$ denotes the outgoing solution and $R_-$ denotes the incoming solution. The angular-dependent part of the action for outgoing solution is usually the same as the incoming solution for rotating black holes, see, e.g., \cite{Li:2008zra,DiCriscienzo:2008dm,Chen:2008vi,Kerner:2008qv,Chen:2008ge,Jiang:2008gq,Vanzo:2011wq,Chen:2014xgj}. Though it is not necessary to be a real function, it will be canceled out when computing the probability of tunneling. However in the present case, equation \eqref{angle} involves the $r$-derivative of the action. A priori, it is necessary to consider $\Phi_+(z, \bz)$ and $\Phi_-(z, \bz)$ as different functions. Nevertheless we will argue that $\Phi_+(z,\bz)$ and $\Phi_-(z,\bz)$ can have real solutions. We will consider the case that the two expressions $\p_z I - g_{vz} \p_r I$ and $\p_{\bz} I - g_{v\bz} \p_r I$ in \eqref{angle} are complex conjugate to each other. The complex conjugate of the first expression for the incoming solution is
\be
\p_{\bz} \left[\int \td r \, \frac{\omega - \sqrt{\omega^2 - \mu^2 g_{vv}}}{-g_{vv}} + \xbar\Phi_-(z,\bz)\right] - g_{v\bz} \frac{\omega - \sqrt{\omega^2 - \mu^2 g_{vv}}}{-g_{vv}} \ ,
\ee
where we have used the fact that $\int \td r \, \frac{\omega - \sqrt{\omega^2 - \mu^2 g_{vv}}}{-g_{vv}}$ is real and a overhead bar on the expression means its complex conjugate. Comparing to the second expression
\be
\p_{\bz} \left[\int \td r \, \frac{\omega - \sqrt{\omega^2 - \mu^2 g_{vv}}}{-g_{vv}} + \Phi_-(z,\bz)\right] - g_{v\bz} \frac{\omega - \sqrt{\omega^2 - \mu^2 g_{vv}}}{-g_{vv}} \ ,
\ee
one obtains that $\xbar\Phi_-(z,\bz)=\Phi_-(z,\bz)$. Hence $\Phi_-(z,\bz)$ is real.

For the outgoing solution, the expression $\int \td r \, \frac{\omega + \sqrt{\omega^2 - \mu^2 g_{vv}}}{-g_{vv}}$ is not real. But the imaginary part of this expression is $4\pi \omega M$ \cite{Chu:2018tzu}, i.e., a constant that will be killed by a $z$ or $\bz$ derivative. Hence, one can show that $\Phi_+(z,\bz)$ is real following the same argument as the incoming case.

The probabilities of crossing the outer horizon each way are respectively given by
\begin{align}
&P_{\text{out}}=\text{exp}\left[-2 \text{Im}I_+\right]= \text{exp}\left[- 8\pi \omega M  - 2 \text{Im}K \right]\ ,\label{out}\\
&P_{\text{in}}=\text{exp}\left[-2 \text{Im}I_-\right]=\text{exp}\left[-2 \text{Im}K\right] \ ,\label{in}
\end{align}
where we set $\hbar$ to unity. {The overall tunnelling probability of a Dirac particle tunneling from inside to outside of the horizon is defined by \cite{Kerner:2007rr}
\be
\Gamma= \frac{P_{\text{out}}}{P_{\text{in}}} \ .
\ee
Inserting \eqref{out} and \eqref{in} yields
\be
\Gamma=\text{exp}\left[- 8\pi \omega M\right] \ ,
\ee
which agrees with the result of scalar particles tunneling from the supertranslated Schwarzschild black hole \cite{Chu:2018tzu}. The Hawking radiation is independent of the supertranslation parameter $f$ and recovers result of Schwarzschild case \cite{Kerner:2007rr}. Hence the supertranslation hair of the Schwarzschild black hole does not affect its Hawking radiation.} Identifying $\Gamma$ with the Boltzmann factor, exp$(-\frac{\omega}{T} )$, we can read off the Hawking
temperature as $\frac{1}{8\pi M}$.

\section{Conclusion}

In this work, we examine the semi-classical property of soft hair implants on black {\cb holes}. Applying the tunneling method, we compute the Hawking radiation of Dirac particles from Schwarzschild black {\cb holes} with linear supertranslation hairs and confirm that the Hawking radiation is not modified by supertranslation hairs. This is another evidence to support the microscopic nature of soft hairs.

\section*{Acknowledgments}

We would like to thank Chong-Sun Chu and Ran Li for very helpful discussions. This work is supported in part by the National Natural Science Foundation of China under Grant No. 11905156, No. 11975164, No. 11935009,
and  Natural Science Foundation of Tianjin under Grant No.~20JCYBJC00910.

\providecommand{\href}[2]{#2}\begingroup\raggedright\endgroup


\begin{thebibliography}{10}

\bibitem{Hawking:2016msc}
S.~W. Hawking, M.~J. Perry, and A.~Strominger, ``{Soft Hair on Black Holes},''
  \href{http://dx.doi.org/10.1103/PhysRevLett.116.231301}{{\em Phys. Rev.
  Lett.} {\bfseries 116} no.~23, (2016) 231301},
  \href{http://arxiv.org/abs/1601.00921}{{\ttfamily arXiv:1601.00921
  [hep-th]}}.

\bibitem{Compere:2016hzt}
G.~Comp\`ere and J.~Long, ``{Classical static final state of collapse with
  supertranslation memory},''
  \href{http://dx.doi.org/10.1088/0264-9381/33/19/195001}{{\em Class. Quant.
  Grav.} {\bfseries 33} no.~19, (2016) 195001},
  \href{http://arxiv.org/abs/1602.05197}{{\ttfamily arXiv:1602.05197 [gr-qc]}}.

\bibitem{Afshar:2016wfy}
H.~Afshar, S.~Detournay, D.~Grumiller, W.~Merbis, A.~Perez, D.~Tempo, and
  R.~Troncoso, ``{Soft Heisenberg hair on black holes in three dimensions},''
  \href{http://dx.doi.org/10.1103/PhysRevD.93.101503}{{\em Phys. Rev. D}
  {\bfseries 93} no.~10, (2016) 101503},
  \href{http://arxiv.org/abs/1603.04824}{{\ttfamily arXiv:1603.04824
  [hep-th]}}.

\bibitem{Mao:2016pwq}
P.~Mao, X.~Wu, and H.~Zhang, ``{Soft hairs on isolated horizon implanted by
  electromagnetic fields},''
  \href{http://dx.doi.org/10.1088/1361-6382/aa59da}{{\em Class. Quant. Grav.}
  {\bfseries 34} no.~5, (2017) 055003},
  \href{http://arxiv.org/abs/1606.03226}{{\ttfamily arXiv:1606.03226
  [hep-th]}}.

\bibitem{Grumiller:2016kcp}
D.~Grumiller, A.~Perez, S.~Prohazka, D.~Tempo, and R.~Troncoso, ``{Higher Spin
  Black Holes with Soft Hair},''
  \href{http://dx.doi.org/10.1007/JHEP10(2016)119}{{\em JHEP} {\bfseries 10}
  (2016) 119}, \href{http://arxiv.org/abs/1607.05360}{{\ttfamily
  arXiv:1607.05360 [hep-th]}}.

\bibitem{Hawking:2016sgy}
S.~W. Hawking, M.~J. Perry, and A.~Strominger, ``{Superrotation Charge and
  Supertranslation Hair on Black Holes},''
  \href{http://dx.doi.org/10.1007/JHEP05(2017)161}{{\em JHEP} {\bfseries 05}
  (2017) 161}, \href{http://arxiv.org/abs/1611.09175}{{\ttfamily
  arXiv:1611.09175 [hep-th]}}.

\bibitem{Chu:2018tzu}
C.-S. Chu and Y.~Koyama, ``{Soft Hair of Dynamical Black Hole and Hawking
  Radiation},'' \href{http://dx.doi.org/10.1007/JHEP04(2018)056}{{\em JHEP}
  {\bfseries 04} (2018) 056}, \href{http://arxiv.org/abs/1801.03658}{{\ttfamily
  arXiv:1801.03658 [hep-th]}}.

\bibitem{Choi:2018oel}
S.~Choi and R.~Akhoury, ``{Soft Photon Hair on Schwarzschild Horizon from a
  Wilson Line Perspective},''
  \href{http://dx.doi.org/10.1007/JHEP12(2018)074}{{\em JHEP} {\bfseries 12}
  (2018) 074}, \href{http://arxiv.org/abs/1809.03467}{{\ttfamily
  arXiv:1809.03467 [hep-th]}}.

\bibitem{Choi:2019fuq}
S.~Choi, S.~Sandeep~Pradhan, and R.~Akhoury, ``{Supertranslation Hair of
  Schwarzschild Black Hole: A Wilson Line Perspective},''
  \href{http://dx.doi.org/10.1007/JHEP01(2020)013}{{\em JHEP} {\bfseries 01}
  (2020) 013}, \href{http://arxiv.org/abs/1910.05882}{{\ttfamily
  arXiv:1910.05882 [hep-th]}}.

\bibitem{Takeuchi:2021rrq}
S.~Takeuchi, ``{A 4D asymptotically flat rotating black hole solution including
  displacement of supertraslation},''
  \href{http://arxiv.org/abs/2102.07363}{{\ttfamily arXiv:2102.07363
  [hep-th]}}.

\bibitem{Strominger:2017aeh}
A.~Strominger, {\em {Black Hole Information Revisited}}.
\newblock 2020.
\newblock \href{http://arxiv.org/abs/1706.07143}{{\ttfamily arXiv:1706.07143
  [hep-th]}}.

\bibitem{Strominger:2017zoo}
A.~Strominger, ``{Lectures on the Infrared Structure of Gravity and Gauge
  Theory},'' \href{http://arxiv.org/abs/1703.05448}{{\ttfamily arXiv:1703.05448
  [hep-th]}}.

\bibitem{Pasterski:2020xvn}
S.~Pasterski and H.~Verlinde, ``{HPS meets AMPS: how soft hair dissolves the
  firewall},'' \href{http://dx.doi.org/10.1007/JHEP09(2021)099}{{\em JHEP}
  {\bfseries 09} (2021) 099}, \href{http://arxiv.org/abs/2012.03850}{{\ttfamily
  arXiv:2012.03850 [hep-th]}}.

\bibitem{Afshar:2016uax}
H.~Afshar, D.~Grumiller, and M.~M. Sheikh-Jabbari, ``{Near horizon soft hair as
  microstates of three dimensional black holes},''
  \href{http://dx.doi.org/10.1103/PhysRevD.96.084032}{{\em Phys. Rev. D}
  {\bfseries 96} no.~8, (2017) 084032},
  \href{http://arxiv.org/abs/1607.00009}{{\ttfamily arXiv:1607.00009
  [hep-th]}}.

\bibitem{Sheikh-Jabbari:2016npa}
M.~M. Sheikh-Jabbari and H.~Yavartanoo, ``{Horizon Fluffs: Near Horizon Soft
  Hairs as Microstates of Generic AdS3 Black Holes},''
  \href{http://dx.doi.org/10.1103/PhysRevD.95.044007}{{\em Phys. Rev. D}
  {\bfseries 95} no.~4, (2017) 044007},
  \href{http://arxiv.org/abs/1608.01293}{{\ttfamily arXiv:1608.01293
  [hep-th]}}.

\bibitem{Javadinezhad:2018urv}
R.~Javadinezhad, U.~Kol, and M.~Porrati, ``{Comments on Lorentz
  Transformations, Dressed Asymptotic States and Hawking Radiation},''
  \href{http://dx.doi.org/10.1007/JHEP01(2019)089}{{\em JHEP} {\bfseries 01}
  (2019) 089}, \href{http://arxiv.org/abs/1808.02987}{{\ttfamily
  arXiv:1808.02987 [hep-th]}}.

\bibitem{Compere:2019rof}
G.~Comp\`ere, J.~Long, and M.~Riegler, ``{Invariance of Unruh and Hawking
  radiation under matter-induced supertranslations},''
  \href{http://dx.doi.org/10.1007/JHEP05(2019)053}{{\em JHEP} {\bfseries 05}
  (2019) 053}, \href{http://arxiv.org/abs/1903.01812}{{\ttfamily
  arXiv:1903.01812 [hep-th]}}.

\bibitem{Lin:2020gva}
F.-L. Lin and S.~Takeuchi, ``{Hawking flux from a black hole with nonlinear
  supertranslation hair},''
  \href{http://dx.doi.org/10.1103/PhysRevD.102.044004}{{\em Phys. Rev. D}
  {\bfseries 102} no.~4, (2020) 044004},
  \href{http://arxiv.org/abs/2004.07474}{{\ttfamily arXiv:2004.07474
  [hep-th]}}.

\bibitem{Takeuchi:2021ibg}
S.~Takeuchi, ``{Hawking flux of 4D Schwarzschild black hole with
  supertransition correction to second-order},'' in {\em {4th International
  Conference on Holography, String Theory and Discrete Approach in Hanoi,
  Vietnam}}.
\newblock 4, 2021.
\newblock \href{http://arxiv.org/abs/2104.05483}{{\ttfamily arXiv:2104.05483
  [hep-th]}}.

\bibitem{Chiang:2020lem}
H.-W. Chiang, Y.-H. Kung, and P.~Chen, ``{Modification to the Hawking
  temperature of a dynamical black hole by a flow-induced supertranslation},''
  \href{http://dx.doi.org/10.1007/JHEP12(2020)089}{{\em JHEP} {\bfseries 12}
  (2020) 089}, \href{http://arxiv.org/abs/2004.05045}{{\ttfamily
  arXiv:2004.05045 [gr-qc]}}.

\bibitem{Wen:2021ahw}
W.-Y. Wen, ``{Dressed tunneling in soft hair},''
  \href{http://dx.doi.org/10.1016/j.physletb.2021.136578}{{\em Phys. Lett. B}
  {\bfseries 820} (2021) 136578},
  \href{http://arxiv.org/abs/2103.00516}{{\ttfamily arXiv:2103.00516
  [hep-th]}}.

\bibitem{Parikh:1999mf}
M.~K. Parikh and F.~Wilczek, ``{Hawking radiation as tunneling},''
  \href{http://dx.doi.org/10.1103/PhysRevLett.85.5042}{{\em Phys. Rev. Lett.}
  {\bfseries 85} (2000) 5042--5045},
  \href{http://arxiv.org/abs/hep-th/9907001}{{\ttfamily arXiv:hep-th/9907001}}.

\bibitem{Kerner:2007rr}
R.~Kerner and R.~B. Mann, ``{Fermions tunnelling from black holes},''
  \href{http://dx.doi.org/10.1088/0264-9381/25/9/095014}{{\em Class. Quant.
  Grav.} {\bfseries 25} (2008) 095014},
  \href{http://arxiv.org/abs/0710.0612}{{\ttfamily arXiv:0710.0612 [hep-th]}}.

\bibitem{Li:2008zra}
R.~Li, J.-R. Ren, and S.-W. Wei, ``{Hawking radiation of Dirac particles via
  tunneling from Kerr black hole},''
  \href{http://dx.doi.org/10.1088/0264-9381/25/12/125016}{{\em Class. Quant.
  Grav.} {\bfseries 25} (2008) 125016},
  \href{http://arxiv.org/abs/0803.1410}{{\ttfamily arXiv:0803.1410 [gr-qc]}}.

\bibitem{DiCriscienzo:2008dm}
R.~Di~Criscienzo and L.~Vanzo, ``{Fermion Tunneling from Dynamical Horizons},''
  \href{http://dx.doi.org/10.1209/0295-5075/82/60001}{{\em EPL} {\bfseries 82}
  no.~6, (2008) 60001}, \href{http://arxiv.org/abs/0803.0435}{{\ttfamily
  arXiv:0803.0435 [hep-th]}}.

\bibitem{Chen:2008vi}
D.-Y. Chen, Q.-Q. Jiang, S.-Z. Yang, and X.-T. Zu, ``{Fermions tunnelling from
  the charged dilatonic black holes},''
  \href{http://dx.doi.org/10.1088/0264-9381/25/20/205022}{{\em Class. Quant.
  Grav.} {\bfseries 25} (2008) 205022},
  \href{http://arxiv.org/abs/0803.3248}{{\ttfamily arXiv:0803.3248 [hep-th]}}.

\bibitem{Kerner:2008qv}
R.~Kerner and R.~B. Mann, ``{Charged Fermions Tunnelling from Kerr-Newman Black
  Holes},'' \href{http://dx.doi.org/10.1016/j.physletb.2008.06.012}{{\em Phys.
  Lett. B} {\bfseries 665} (2008) 277--283},
  \href{http://arxiv.org/abs/0803.2246}{{\ttfamily arXiv:0803.2246 [hep-th]}}.

\bibitem{Chen:2008ge}
D.-Y. Chen, Q.-Q. Jiang, and X.-T. Zu, ``{Hawking radiation of Dirac particles
  via tunnelling from rotating black holes in de Sitter spaces},''
  \href{http://dx.doi.org/10.1016/j.physletb.2008.05.064}{{\em Phys. Lett. B}
  {\bfseries 665} (2008) 106--110},
  \href{http://arxiv.org/abs/0804.0131}{{\ttfamily arXiv:0804.0131 [hep-th]}}.

\bibitem{Jiang:2008gq}
Q.-Q. Jiang, ``{Dirac particles' tunnelling from black rings},''
  \href{http://dx.doi.org/10.1103/PhysRevD.78.044009}{{\em Phys. Rev. D}
  {\bfseries 78} (2008) 044009},
  \href{http://arxiv.org/abs/0807.1358}{{\ttfamily arXiv:0807.1358 [hep-th]}}.

\bibitem{Vanzo:2011wq}
L.~Vanzo, G.~Acquaviva, and R.~Di~Criscienzo, ``{Tunnelling Methods and
  Hawking's radiation: achievements and prospects},''
  \href{http://dx.doi.org/10.1088/0264-9381/28/18/183001}{{\em Class. Quant.
  Grav.} {\bfseries 28} (2011) 183001},
  \href{http://arxiv.org/abs/1106.4153}{{\ttfamily arXiv:1106.4153 [gr-qc]}}.

\bibitem{Chen:2014xgj}
D.~Chen, H.~Wu, H.~Yang, and S.~Yang, ``{Effects of quantum gravity on black
  holes},'' \href{http://dx.doi.org/10.1142/S0217751X14300543}{{\em Int. J.
  Mod. Phys. A} {\bfseries 29} no.~26, (2014) 1430054},
  \href{http://arxiv.org/abs/1410.5071}{{\ttfamily arXiv:1410.5071 [gr-qc]}}.

\bibitem{Angheben:2005rm}
M.~Angheben, M.~Nadalini, L.~Vanzo, and S.~Zerbini, ``{Hawking radiation as
  tunneling for extremal and rotating black holes},''
  \href{http://dx.doi.org/10.1088/1126-6708/2005/05/014}{{\em JHEP} {\bfseries
  05} (2005) 014}, \href{http://arxiv.org/abs/hep-th/0503081}{{\ttfamily
  arXiv:hep-th/0503081}}.

\end{thebibliography}
\end{document}